\documentclass[prb,aps,amsmath,amssymb,twocolumn,showpacs,floatfix]{revtex4}
\usepackage[dvips]{graphicx}
\usepackage{bm}

\def\be{\begin{equation}}
\def\ee{\end{equation}}
\def\bea{\begin{eqnarray}}
\def\eea{\end{eqnarray}}

\newcommand\br{{\bm r}}

\newcommand\bn{{\bm n}}

\newcommand\bE{{\bm E}}
\newcommand\bme{{\bm e}}

\newcommand\tq{{\tau_{\rm q}}}
\newcommand\ttr{{\tau_{\rm tr}}}
\newcommand\tsh{{\tau_{\rm sh}}}
\newcommand\tsm{{\tau_{\rm sm}}}
\newcommand\tpi{{\tau_{\rm \pi}}}
\newcommand\tin{{\tau_{\rm in}}}
\newcommand\wc{{\omega_c}}
\newcommand\ve{\varepsilon}

\newcommand\w{\omega}
\newcommand\St{{\rm St}}

\renewcommand\j{{\bm j}}
\newcommand\bj{{\bm j}}

\newcommand\bzeta{\bm{\zeta}}

\begin{document}

\title{Mechanisms of the microwave photoconductivity in 2D electron systems with mixed disorder}
\author{I.A.~Dmitriev$^{1,2,*}$}
\author{M.~Khodas$^{3}$}
\author{A.D.~Mirlin$^{1,2,\#}$}
\author{D.G.~Polyakov$^{1}$}
\author{M.G.~Vavilov$^{4}$}
\affiliation{$^{1}$Institut f\"ur Nanotechnologie, Forschungszentrum
Karlsruhe, 76021 Karlsruhe, Germany\\
$^{2}$Institut f\"ur Theorie der kondensierten Materie,
Universit\"at Karlsruhe, 76128 Karlsruhe, Germany}
\affiliation{$^{3}$Department of Condensed Matter Physics and Materials
Science, Brookhaven National Laboratory, Upton, NY 11973-5000, USA}
\affiliation{$^{4}$Department of Physics, University of Wisconsin, Madison, WI 53706,
USA }
\date{\today}

\begin{abstract}\noindent
We present a systematic study of the microwave-induced oscillations in the 
 magnetoresistance of a 2D electron gas for mixed disorder including both short-range and long-range components.
 The obtained photoconductivity tensor contains contributions of four distinct transport 
 mechanisms. We show that the photoresponse depends crucially on the relative weight of the short-range component of disorder. 
Depending on the properties of disorder, the theory allows 
 one to identify the temperature range within which the photoresponse is dominated by one 
 of the mechanisms analyzed in the paper.

\end{abstract}

\maketitle

\section{Introduction}\label{s1}
\noindent
During the last
decade many experimental and theoretical advances have been made
in the field of nonequilibrium magnetotransport in 2D electron
systems. The research activity was triggered by the discovery of a
number of beautiful nonequilibrium phenomena, 
the microwave-induced resistance oscillations
(MIRO)\cite{zudov01,ye01} in the first place, governed by the ratio $\omega/\wc$ of
the radiation frequency and the cyclotron frequency $\wc=|e|B/mc$.
Further experiments on the
MIRO\cite{mani02}$^{\!-\!}$\cite{highFdu}
led to a spectacular
observation of zero resistance states (ZRS)\cite{mani02,zudov03,yang03,dorozhkin03,willett03} which were later
explained as a result of instability leading to the formation of
current domains.\cite{AAM03}

These discoveries stimulated an intense theoretical research\cite{AAM03}$^{\!-\!}$\cite{DDM09}
which has led to significant advances in understanding the nonequilibrium transport in high Landau levels 
(LLs). Initially, the MIRO were attributed to the ``{\it displacement}'' mechanism which accounts for spatial displacements
of semiclassical electron orbits assosiated with the radiation-assisted
scattering off disorder.\cite{DSRG03,ryzhii70,VA04} The preferred direction of such
displacements along the symmetry-breaking dc field
oscillates as a function of $\omega/\wc$ due to a periodic modulation in the density
of states (DOS) $\nu(\ve)=\nu(\ve+\hbar\wc)$ in a magnetic field. 
The displacement mechanism results in temperature
independent MIRO with the phase and period as those 
observed in the experiments.

The dominant contribution to the MIRO was later shown to be due to the
``{\it inelastic}'' mechanism associated with a radiation-induced
changes in the energy distribution
of electrons.\cite{DMP03,DVAMP0405} The nonequilibrium part of the
distribution oscillates with energy due to the oscillatory DOS and its
amplitude is
controlled by the inelastic relaxation time $\tin\propto T^{-2}$.\cite{DVAMP0405}
As a consequence, the inelastic contribution to
the MIRO
provided an explanation of the observed strong temperature dependence of the MIRO.

Further, a recent systematic study of the photoresponse
revealed two additional mechanisms of the MIRO, 
``{\it quadrupole}'' and ``{\it photovoltaic}.''\cite{DMP07} In the
quadrupole mechanism, the microwave radiation leads to excitation
of the second angular harmonic of the distribution function. The
dc response in the resulting nonequilibrium state yields an
oscillatory contribution to the Hall part of the photoconductivity
tensor which violates Onsager symmetry.\cite{onsager} In the photovoltaic
mechanism, a combined action of the microwave and dc fields
produces non-zero temporal harmonics of the 
distribution function. The response of this state to an ac electric field
contributes to both the longitudinal and Hall dc photoconductivity.

The theory developed in Refs.~\onlinecite{VA04,DVAMP0405,DMP07} assumed that the only relevant source of
disorder potential is that created by remote ionized donors which are separated
from the plane of 2D electrons by a wide spacer of width strongly exceeding the
Fermi wavelength, $d\gg k_F^{-1}$. Such smooth disorder is characterized by
scattering on small angles $\sim 1/(k_F d)$, resulting in the long transport scattering
time, $\ttr\sim\tq(k_F d)^{2}\gg\tq$, and the
backscattering time, $\tpi\propto\tq \exp(2 k_F d)$;
here $\tq$ is the total disorder-induced quantum scattering time.
For a smooth disorder potential, the inelastic contribution dominates over
the displacement contribution for $\tau_{\rm in}>\tau_q$, the condition which is
fulfilled in the whole temperature range where the MIRO were observed in the early
experiments.\cite{zudov01,ye01,mani02,zudov03,yang03,dorozhkin03}
 The quadrupole and photovoltaic mechanisms are the only ones
yielding oscillatory corrections to the Hall part of the photoconductivity
tensor, with the magnitude comparable to that of
the displacement contribution to the diagonal part.

In addition to the MIRO, 2D electrons exhibit
nonequilibrium magnetooscillations when a strong current is applied to a sample in the absence
of the microwave radiation.\cite{HIRO,Bykov05,inelasticDC,strongDC,zudov09dc} These oscillations in the nonlinear resistivity,
called hall-induced resistance oscillations (HIRO), were attributed to the
resonant backscattering off disorder between LLs tilted by the
electric field.\cite{HIRO,VAG07} Therefore, in contrast to the MIRO, the HIRO require a sufficient amount
of short-range disorder in order to be visible in experiment. 
This disorder potential is created by residual charged impurities
in the vicinity of the 2D electron plane. Being inevitably
present in quantum Hall structures, such impurities are believed to limit
the mobility of ultra-high mobility samples.
Recent experimental\cite{ACDC,acdc2,fracACDC} and
theoretical\cite{KV08} study of the interplay of the MIRO and HIRO demonstrated that the
displacement contribution to the MIRO is very sensitive to the nature of
disorder. In fact, it is known that various transport phenomena in
2D electron systems crucially depend on the type of disorder, 
including quantum magnetooscillations in the dc and ac transport, momentum-dependent conductivity, 
quasiclassical memory effects, commensurability oscillations in lateral superlattices, and interaction-induced quantum magnetoresistance (for a review, see Ref.~\onlinecite{PSS08}). 
Therefore, generalization of the previous studies on the MIRO to
the case of a generic disorder potential is highly desirable.

Here we present a comprehensive analysis of the photoresponse, taking
into account all four contributions to
the MIRO\cite{DMP07} and using a generic-disorder model
which is characterized by an arbitrary dependence of the elastic scattering rate
\be\label{tt}
\tau_\theta^{-1}=\sum_{n=-\infty}^\infty \tau_n^{-1}
e^{in\theta},\quad \tau_n=\tau_{-n}
\ee
on the scattering angle $\theta$. We analyze the obtained results 
for a realistic model of mixed disorder formed by superposition of a smooth random potential
and short-range scatterers.
The theory allows one to identify the temperature range
where a specific mechanism dominates the photoresponse for different types of
 disorder.

The paper is organized as follows. In Sec.~\ref{s2} we present a kinetic equation approach to the problem.
In Sec.~\ref{s3} we obtain the photoconductivity tensor which is analyzed in Sec.~\ref{s4} for
different types of disorder. In Sec.~\ref{s5} we disccuss possible temperature
regimes of the MIRO taking into account the interaction-induced broadening of Landau levels.
Main findings are summarized in  Sec.~\ref{s6}.

\section{Formalism}\label{s2}
\noindent

To begin with, we briefly outline the main steps in setting up the formalism. 
We consider 2D electrons in a classically strong magnetic field $B$ in
the presence of a random potential, a dc electric field,
\be
\bE_{\rm dc}=(E_x,E_y),
\ee
and a microwave field
\be
\bE_\w(t)=E_\omega \sum_{\pm} s_\pm {\rm Re}\left[\,{\bme}_\pm e^{i\w t}\right],
\ee
where $\sqrt{2}{\bme}_\pm={\bme}_x\pm i{\bme}_y$, and $s_\pm$
parameterize polarization of the microwaves ($s_+^2+s_-^2=1$).
The main parameters in the problem are related to each other as follows:
\[\ve_F\gg T\,,\,\omega\,,\,\omega_c\,,\,\tq^{-1} \gg\ttr^{-1},\]
where
$\ve_F$ is the Fermi energy and we use $k_{B}=\hbar=1$.
We adopt the approach \cite{VA04,DVAMP0405,DMP07,KV08} to the problem,
based on the quantum Boltzmann
equation (QBE) for the semiclassical distribution function $f(\ve,\varphi,t)$ at
higher LLs,
\be\label{QBE}
(\partial_t+\wc\partial_\varphi)f+\tau_{\rm in}^{-1}(\langle f \rangle
-f_T)=\St\{f\}\,.
\ee
In the inelastic collision integral $\propto\tau_{\rm in}^{-1}$, $f_T(\ve)$ is the Fermi-Dirac distribution. The angular
brackets $\langle\dots\rangle$ denote
averaging over the direction  $\varphi$ of the kinematic momentum.

The QBE (\ref{QBE}) allows us to
treat the interplay of the disorder, the
Landau quantization, and the external fields, which are all
included into the impurity collision integral $\St\{f\}$.
The field-dependent collision integral appears as a
consequence of transition to a moving coordinate frame, $\br\to\br-\bzeta_t$, where
\be\label{bz}
\partial_t \bzeta_t=
\left(\frac{\partial_t -\w_c \hat{\varepsilon}}{\partial_t^2+\w_c^2}\right)
\frac{e}{m}[\bE_{\rm dc}+\bE_\w(t)]
\ee
and $\hat{\varepsilon}_{xy}=-\hat{\varepsilon}_{yx}=1$.
In the new frame, the electric field is absent,
but the impurities are moving and can change the energy of electrons.
The effects of generic
disorder and external fields are encoded in the integral operator\cite{KV08}
\be \label{K}
\hat{\mathcal{K}}_{t_2,t_1;\varphi}\{F(\varphi)\} =
\int \frac{ d \varphi' }{ 2 \pi}
\frac{ e^{ i p_{\rm F}
(\bn_{\varphi}-\bn_{\varphi'})(\bzeta_{t_2}-\bzeta_{t_1})}}{
\tau_{\varphi - \varphi'} } F(\varphi')\,
\ee
where $p_F=m v_F$ is the Fermi momentum, the unit vector $\bn_{\varphi}=(\cos\varphi, \sin\varphi)$, and $F(\varphi)$ is an arbitrary function.
In terms of $\hat{\cal K}$, the scattering integral in the time representation is given by
the expression
\begin{equation}\begin{split}
\St\{f\}_{t_2,t_1}&=\int\! dt_3\, \left[\,\hat{\cal K}_{t_2,t_1}\{g_{t_2-t_3}^R
f_{t_3,t_1}-f_{t_2,t_3}g_{t_3-t_1}^A\}\right.
 \\
-&\left. f_{t_3,t_1}\hat{\cal  K}_{t_2,t_3}\{g_{t_2-t_3}^R\}
+f_{t_2,t_3}\hat{\cal K}_{t_3,t_1}\{g_{t_3-t_1}^A\}\,\right]~.
\label{St}
\end{split}
\end{equation}

We are going to explore the regime of overlapping LLs, $\wc\tq\ll 1$, when the spectral 
functions
\be
g^{R}_{t_2-t_1}=[g^{A}_{t_2-t_1}]^\dagger=\delta(t_2-t_1)/2-\lambda\delta(t_2-t_1-t_B),
\ee
depend on a single parameter
\be\label{lambda}
\lambda=\exp(-\pi/\wc \tq)\ll 1
\ee
and are insensitive to
the external electric fields. The corresponding density of states (DOS) is
\be\label{DOS}
\nu(\ve)=\nu_0[1-2\lambda\cos(\ve t_B)],
\ee
where $t_B=2\pi/|\wc|$
is the cyclotron period and $\nu_0=m/2\pi$ is the DOS at $B=0$ per spin degree of freedom.

The Wigner transform (Fourier transform with respect to
$t_-=t_2-t_1$) of the scattering integral, Eq.~(\ref{St}), defines
the QBE for the distribution function $f(\ve,\varphi,t)$, which
can be approximated by
\be
\label{f}
f=f_T+(\partial_\ve
f_T)\{\phi_0^D-2\lambda {\rm Re}[\phi_1\exp(i\ve
t_B)]+2\lambda^2\phi_0^{(2)}\},
\ee
where
\be
\phi_0^D(\varphi)=e v_F \bE_{\rm dc}\cdot\bn_{\varphi}/\wc^2\ttr
\ee
is the classical
part leading to the Drude expression for the current (\ref{drude}), and $t=(t_2+t_1)/2$. By the
symmetry of the kernel (\ref{K}),
\be
\phi_\bot\equiv{\rm
Im}\phi_1(\varphi,t)
\ee
includes even angular
harmonics of the distribution
function and is governed by the real part of the kernel, Eq.~(\ref{K}):
\bea
\nonumber &&(\partial_t+\wc\partial_\varphi)\phi_\bot
+\tau_{\rm in}^{-1}\left\langle\phi_\bot\right\rangle\\
\label{kinbot}
&&\left. \qquad =(\frac{1}{2}\partial_t-\partial_{t_-}+\phi_\bot)
{\rm Re}\hat{\cal K}_{t_2,t_1}\{1\}\right|_{t_2=t_1+t_B}~.
\eea

The dc electric current,
\be
\label{curgen}
{\j}=2ev_F\int d\ve \,\nu(\ve) \langle
\overline{{\bn}_\varphi f(\ve,\varphi,t)}
\rangle- 2 e \nu_0 \ve_F\overline{\partial_t \bzeta(t)}~,
\ee
is determined by the first angular harmonic of the distribution
function, Eq.~(\ref{f}), which is present in ${\rm Re}\phi_1$, $\phi_0^{(D)}$, and  $\phi_0^{(2)}$.
The bar denotes time averaging over the period
of the microwave field in the
steady state.

Our aim is to calculate the current (\ref{curgen})
for generic disorder (\ref{tt}) to order $E_\w^2 E_{dc}$
 for temperatures $T\gg\w,\wc$.
This regime is the most relevant to experiments on the MIRO and allows for a
reliable comparison between the theory and experiment.
In the absence of quantum corrections ($\lambda=0$) the current is
given by the Drude formula,
\be\label{drude}
{\bj}^D=2\sigma_D (1
-\wc\ttr\hat{\varepsilon})\bE_{\rm dc},
\ee where $\sigma_D=e^2\nu_0
v_F^2/2\wc^2\ttr$. Neither strong dc field nor microwaves
modify this result as long as $\lambda=0$ and the energy
dependence of elastic scattering (weak at $\wc\ll\ve_F$) is
neglected.\cite{classical}
All quantum corrections to $\nu(\ve) f(\ve)$ of
first order in $\lambda$ oscillate with energy
and are exponentially suppressed at $2\pi^2 T/\wc\gg 1$
similar to Shubnikov-de Haas oscillations.
The leading quantum corrections
which survive the temperature smearing at $T\gg\wc$ are of order
$\lambda^2$ and can be written using Eqs.~(\ref{f}), (\ref{curgen}) as
\be\label{j}
{\bf j}={\bf
j}^D-4\lambda^2 ev_F\nu_0 ({\rm Re}J,{\rm Im}J)^{\rm T}
\ee
with
\bea\nonumber &&J=\left\langle
\frac{e^{i\varphi}}{\wc}{\left(\,\overline{\partial_{t_-}\hat{\mathcal{K}}_{t_2,t_1}\{1\}}
-\overline{\phi_\bot\hat{\mathcal{K}}_{t_2,t_1}\{1\}}\right.}\right.\\
&&\phantom{aaaaaaaaaaaaa}-\left.\left.\overline{\hat{\mathcal{K}}_{t_2,t_1}\{\phi_\bot\})}\,\right)\right\rangle_{t_-=t_B}~.
\label{J}
\eea

\section{Photoconductivity tensor}\label{s3}
\noindent
We show that the current $J$ [Eq.~\eqref{J}] depends in an essential way on the properties of disorder; specifically, on whether it scatters isotropically (short-range disorder) or primarily on small angles (long-range disorder). A straightforward calculation of $J$ to the first order in $E_{dc}$ and to the second order in $E_\omega$ using
the solution $\phi_\bot$ of Eq.~(\ref{kinbot}) gives the following expression for
the current (\ref{j}):
\bea\label{cur}
&&{\bj}={\bj}_D+4 \sigma_D \lambda^2\bE_{\rm dc}+\hat{\sigma}^{\rm (ph)}\bE_{\rm dc}~,\\
\label{matrix} &&\hat{\sigma}^{\rm (ph)}=-4 \sigma_D\left(\!
\begin{array}{cc}d_s+d_a & h_s+h_a \\ h_s-h_a & d_s-d_a \end{array}
\!\right).
\eea
The terms of different symmetry $d_{s,a}$ and $h_{s,a}$ in the photoconductivity tensor $\hat{\sigma}^{\rm (ph)}$ 
are given below for arbitrary strength of the short-range and smooth components of disorder. The diagonal part of the photoconductivity tensor consists of the isotropic, $d_s$, and anisotropic, $d_a$, components,
where
\begin{subequations}\label{dss}
\bea\label{dsdef}
d_s & = & d_s^{(A)}+d_s^{(B)}+d_s^{(D)},\\
\label{dsA}
d_s^{(A)}&=&  \lambda^2 [\sin^2 w + w\sin
2w]\frac{\ttr}{2\tau_*}\sum_\pm{\cal E}_\pm^2~,\\
\label{dsB}
d_s^{(B)}&=& \lambda^2 [w\sin 2w] \frac{2\tin}{\ttr}\sum_\pm{\cal
E}_\pm^2~,\\
\nonumber
d_s^{(D)}&=&-\lambda^2 [w\sin^2 w]
\\\label{dsD}&\times&\sum_\pm{\cal
E}_\pm^2\left(\frac{2}{\w\ttr}+\frac{2\ttr\tilde{\tau}^{-2}}{\w\pm2\wc}\right)
\eea
and
\bea
\label{dadef}
d_a & = & d_a^{(A)}+d_a^{(D)},\\
\label{daA}
d_a^{(A)}&=&- \lambda^2 [\sin^2 w + w\sin 2w]\frac{\ttr}{2\tau_*}{\cal E}_+{\cal E}_-~,
\\\label{daD}
d_a^{(D)}&=& \lambda^2 [w\sin^2 w]\frac{4}{\w\ttr}{\cal E}_+{\cal E}_-~.
\eea
\end{subequations}
In the off-diagonal part, $h_a$ represents the microwave--induced correction to
the Hall conductivity, while $h_s$ is the anisotropic contribution
to the dissipative conductivity (violating the Onsager symmetry)\cite{DMP07, onsager}
\begin{subequations}\label{hss}
\bea\label{hsC}
h_s=h_s^{(C)}&=&-\lambda^2 [w\sin 2w](\ttr/\wc\tilde{\tau}^2){\cal E}_+{\cal E}_-~,
\\\nonumber
h_a=h_a^{(D)}&=&-\lambda^2 [2\sin^2 w + w\sin
2w]\\\label{haD}&\times&\sum_\pm\pm{\cal
E}_\pm^2\left(\frac{1}{\w\ttr}+\frac{\ttr\tilde{\tau}^{-2}}{\w\pm2\wc}\right).
\eea
\end{subequations}
All microwave-induced corrections (\ref{dss}) and (\ref{hss})
show oscillations with the ratio
\be\label{w}
w=\frac{\omega t_B}{2}=\frac{\pi\omega}{|\wc|}
\ee
and are of the second order in
the microwave field $E_\w$ measured in dimensionless
units
\be\label{Epm}
{\cal E}_\pm=s_\pm \frac{e v_F E_\w}{\w(\w\pm\wc)}\,.
\ee

In Eqs.~(\ref{dss}) and (\ref{hss}), the superscripts denote the contributions from the displacement (A),
inelastic (B), quadrupole (C) and photovoltaic (D) mechanisms.\cite{DMP07}
The displacement contribution, (A), comes from the first term in Eq.~(\ref{J}). In the absense of the microwave
radiation, $E_\w=0$, this term produces the quantum correction to the Drude conductivity given by the second term in Eq.~(\ref{cur}).
At order $E_{\rm dc} E_\w^2$, the first term in Eq.~(\ref{J}) accounts for the microwave-assisted
 disorder scattering in the presence of the symmetry-breaking dc field. Therefore, in the
displacement mechanism the radiation directly affects the first angular
harmonic of the distribution function, and thus the dc current, see Eqs.~(\ref{dsA}) and (\ref{daA}).
Mechanisms (B), (C), and (D) are related to the microwave excitation of
even angular harmonics of the distribution function governed by Eq.~(\ref{kinbot}).
Their contribution to the current appears at the same order $E_{\rm dc} E_\w^2$ and is described by two last terms in Eq.~(\ref{J}).
The microwave-induced changes in the time-independent zero angular harmonics of the distribution function
are controlled by the inelastic relaxation, and lead to the inelastic contribution, (B), Eq.~(\ref{dsB}).
In the quadrupole mechanism, (C), the microwave radiation excites the second angular
harmonic of the distribution function, $\phi_\bot^{(C)}\propto E_\w^2\cos 2\phi$,
which also contributes to the dc current at order $E_{\rm dc}E_\w^2$ after
substitution into Eq.~(\ref{J}) with ${\mathcal{K}}\propto E_{\rm dc}$, see Eq.~(\ref{hsC}).
In the photovoltaic mechanism, (D), a combined action of the
microwave and dc fields in Eq.~(\ref{kinbot}) produces both zero and second angular harmonics
of the distribution function $\phi_\bot$ which oscillate in time with the microwave frequency.
The ac response in the resulting state
[Eq.~(\ref{J}) with ${\mathcal{K}}\propto E_\w$ and $\phi_\bot\propto E_\w E_{\rm dc}$]
contributes to the
longitudinal and Hall parts of $\hat{\sigma}^{\rm (ph)}$, see Eqs.~(\ref{dsD}), (\ref{daD}), and (\ref{haD}).

The photoconductivity tensor (\ref{matrix}) depends on the inelastic scattering rate
 $\tau_{\rm in}^{-1}$
which enters the inelastic contribution (\ref{dsB}) and on
four different disorder scattering rates.
The latter are expressed in terms of the angular harmonics $\tau_n$ of the disorder scattering rate, Eq.~(\ref{tt}), as
\begin{subequations}\label{tau}
\bea
\label{tq}
\frac{1}{\tq}&=&\frac{1}{\tau_0},\\
\label{ttr}
\frac{1}{\ttr}&=&\frac{1}{\tau_0}-\frac{1}{\tau_1},\\
\label{tstar}
\frac{1}{\tau_*}&=&\frac{3}{\tau_0}-\frac{4}{\tau_1}+\frac{1}{\tau_2},\\
\label{ttilde}
\frac{1}{\tilde{\tau}}&=&\frac{1}{\tau_1}-\frac{1}{2\tau_0}-\frac{1}{2\tau_2}.
\eea
\end{subequations}
We emphasize that as compared to the case of smooth disorder, where the
rates $\tau_{\rm q}^{-1}$ and $\tau_{\rm tr}^{-1}$ fully parametrize the
photoconductivity,\cite{DMP07} two additional rates $\tau_*^{-1}$ and $\tilde\tau^{-1}$
are required in general situation.
The rate $\tau_*^{-1}$ controls the magnitude of the displacement contributions \eqref{dsA} and \eqref{daA}, reproducing the result of Ref.~\onlinecite{KV08}. Note that, in view of $\sigma_D\propto\tau_{\rm tr}^{-1}$, the product $\sigma_D d_s^{(A)}\propto\tau_*^{-1}$ entering Eq.~\eqref{matrix} is actually independent of $\ttr$. The rate $\tilde{\tau}^{-1}$ enters the quadrupole and photovoltaic contributions \eqref{dsD}, \eqref{daD}-\eqref{haD}.
The relation between the scattering rates (\ref{tau}) strongly depends
on the type of disorder potential in the plane of 2D electrons (short-range vs. long-range).
We discuss this dependence  in the following section for a realistic
model of the disorder.

\section{Dependence of MIRO on type of disorder}\label{s4}
\noindent
In this section we study the relative magnitude of the contributions of
mechanisms (A)-(D) for different types of disorder. Using Eqs.~\eqref{dss}, \eqref{hss} and \eqref{tau}
we demonstrate that the MIRO amplitude is
very sensitive to the details of the disorder potential and may provide a valuable information about
the disorder which can not be extracted from the conventional transport measurements
of the mobility and the Shubnikov-de Haas (SdH) oscillations.\cite{noteSdH}

\subsection{Mixed-disorder model}\label{ss41}
The major source of elastic scattering in high-mobility Hall
structures is a smooth random potential created by remote donors
that are separated by a large spacer of width $d\gg k_F^{-1}$
from the 2DEG plane. The disorder model which takes into account only
such remote charged impurities was used in a number of theoretical works on the MIRO\cite{VA04,DVAMP0405,DMP07}
since it allows one to account for the experimentally relevant small--angle scattering
condition $\tau_{\rm q}\ll\tau_{\rm tr}$ and to consider in an unambiguous way
the region of magnetic fields $\tau_{\rm tr}^{-1}\ll\wc\lesssim \tau_{\rm
q}^{-1}$. Here we implement a more realistic\cite{mixed_exp} ``two--component'' (or ``mixed'') model of
disorder\cite{mirlin01} which includes strong short-range scatterers in addition to
the smooth potential.

In terms of the angular harmonics of the elastic scattering rate (\ref{tt}), the mixed disorder model is formulated as\cite{VAG07,KV08}
\begin{align} \label{mixed-disorder}
\frac{ 1 }{ \tau_n } = \frac{ \delta_{n,0} }{ \tsh } +\frac{ 1
}{ \tsm }\frac{ 1 }{ 1 + \chi n^2} \, .
\end{align}
Here the  isotropic scattering rate $\tau_{\rm sh}^{-1}$ characterizes
the short-range disorder created by residual impurities
located at or near the interface, while the smooth part is
defined by the total (quantum) rate $\tau_{\rm sm}^{-1}$ and by the parameter
$\chi=(k_F d)^{-2}\ll 1$ giving a characteristic
scattering angle $\theta\sim\sqrt{\chi}\ll 1$.

Since in high-mobility samples $\ttr\gg\tq$, we imply that $ \tau_{\rm sh}^{-1}\ll\tau_{\rm sm}^{-1}$
and  the smooth component dominates in
the quantum relaxation rate
\be\label{tqmix}
\frac{1}{\tq}=\frac{1}{\tau_0}=\frac{1}{\tsh}+\frac{1}{\tsm}.
\ee
At the same time, the relative weight of the sharp and smooth components in the
transport relaxation rate
\be\label{ttrmix}
\frac{1}{\ttr}=\frac{1}{\tsh}+\frac{1}{\tsm} \frac{\chi}{ 1 + \chi}
\ee
can be arbitrary.

\subsection{Photoconductivity for smooth disorder}\label{ss42}\noindent
It is useful to recall the results of Ref.~\onlinecite{DMP07} for the case of smooth disorder and to see how they are obtained from Eqs.~\eqref{dss},
\eqref{hss} and \eqref{tau}. Putting $\tau_{\rm sh}^{-1}=0$ (no short-range disorder), the scattering rates that enter Eq.~\eqref{tau} can be rewritten as
\begin{subequations}\label{tausmooth}
\bea
\label{tqsmooth}
\frac{1}{\tq}&=&\frac{1}{\tsm},\\
\label{ttrsmooth}
\frac{1}{\ttr}&=&\frac{\chi}{\tq},\\
\label{tstarsmooth}
\frac{1}{\tau_*}&=&\frac{12\chi^2}{\tq},\\
\label{ttildesmooth}
\frac{1}{\tilde{\tau}}&=&\frac{1}{\ttr}.
\eea
\end{subequations}

As a result, in the case of smooth disorder the ratio of the inelastic
contribution \eqref{dsB} to the displacement contribution \eqref{dsA}  is
\be\label{BtoA}
\frac{d_s^{(B)}}{d_s^{(A)}}=\frac{4\tin\tau_*}{\tau_{\rm tr}^2}=\frac{\tin}{3\tq}, \quad\tau_{\rm sh}^{-1}=0.
\ee
This relation follows from Eqs.~\eqref{dsA} and \eqref{dsB} if one disregards the term $\propto \sin^2 w$ which is small at $w=\pi\w/\wc\gg 1$.

For typical conditions of the MIRO experiments, the parameters entering equations Eqs.~\eqref{dss},
\eqref{hss} and \eqref{tausmooth} can be estimated as $\wc\sim\w\sim\tq^{-1}$ and $\chi=\tq/\ttr\simeq10^{-2}$.
Using the estimate $\tin\sim\ve_F/T^2$ for the inelastic scattering time, we obtain
$\tin\sim\ttr$ at $T=1K$. Therefore, the ratio \eqref{BtoA} remains small up to $T\sim 10K$ (where the MIRO are strongly suppressed).
The magnitude of the oscillations produced by the quadrupole and photovolatic mechanisms is of the order of the displacement contribution,
as they contain a small prefactor $(\w\ttr)^{-1}\sim(\wc\ttr)^{-1}\sim\tq/\ttr$.

We come to the conclusion that, in the case of smooth disorder, the photoconductivity tensor
is dominated by inelastic mechanism (B). It produces the isotropic diagonal contribution \eqref{dsB} proportional to $\tin\sim\ve_F/T^2$.
Other mechanisms produce weak anisotropy of the photoconductivity tensor [Eqs.~\eqref{daA}, \eqref{daD}, \eqref{hsC}] and
govern the Hall part of the photoconductivity, \eqref{haD}.

\subsection{Photoconductivity for mixed disorder}\label{ss43}
Now we return to the mixed-disorder model \eqref{mixed-disorder} with nonzero
$\tau_{\rm sh}^{-1}$ and analyze 
the dependence of various contributions to the photoconductivity tensor, Eq.~(\ref{matrix}),
on the
weight $x =\ttr/\tsh$ of the sharp
component of disorder in the transport relaxation rate \eqref{ttrmix}. In this analysis, we fix both $\ttr$, Eq.~\eqref{ttrmix},  and $\tq$, Eq.~\eqref{tqmix},
and use as a parameter the ratio $\alpha=\ttr/\tq$ which can be extracted from the experiment.\cite{noteSdH}

Inspection of Eqs.~\eqref{dss}, \eqref{hss} shows that only the inelastic contribution \eqref{dsB} is independent of the type of disorder at fixed  $\ttr$  and $\tq$,
\be\label{const}
d_s^{(B)}\propto\tin/\ttr={\rm const}(x)\,.
\ee

\begin{figure}[ht]
\centerline{
\includegraphics[width=0.95\columnwidth]{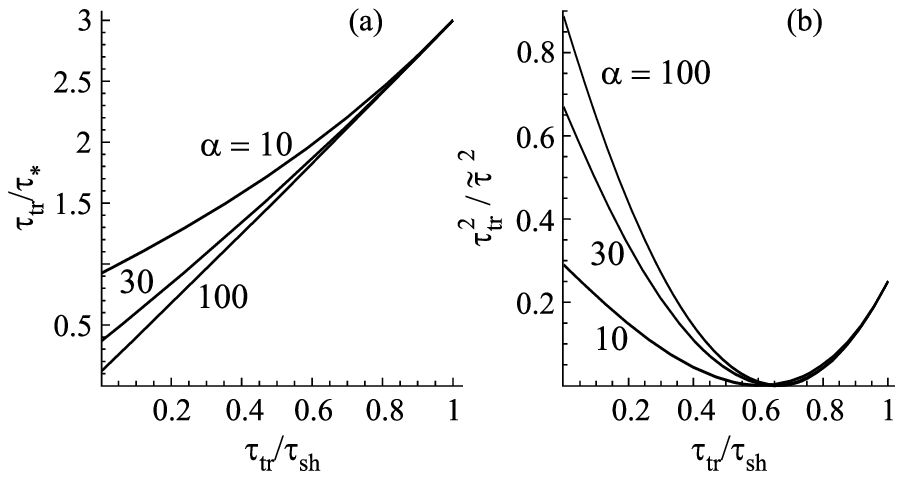}}
\caption{Dependence of the relaxation rates $\tau_{*}^{-1}$ (a) and $\ttr/\tilde{\tau}^2$ (b) on the type of disorder (parametrized by the weight $\ttr/\tsh$ of the short-range component of disorder in the transport relaxation rate) for fixed $\tau_{\rm tr}$ and $\tau_{\rm q}$ for three values of $\alpha=100,30,10$. 
}
 \label{fig1}
 \end{figure}

The displacement contributions $d_s^{(A)}$ and $d_a^{(A)}$, Eqs.~\eqref{dsA} and \eqref{daA}, are proportinal to the ratio 
\be\label{star}
\frac{\ttr}{\tau_{*}}=3\,\frac{\alpha x-5 x+4}{\alpha-4 x+3}\,,
\ee
which is illustrated in Fig.~\ref{fig1}a for several values of the parameter $\alpha=\ttr/\tq$.
Since in high-mobility structures the parameter $\alpha$ is always large, Eq.~\eqref{star} demonstrates that the displacement contribution is parametrically enhanced in the case
of short-range disorder, $\tau_*|_{x=1}/\tau_*|_{x=0}=4\alpha\gg 1$. 

For the ratio $(\ttr/\tilde{\tau})^2$ which enters the photovoltaic and quadrupole contributions
\eqref{dsD}-\eqref{haD}, we obtain the dependence illustrated in Fig.~\ref{fig1}b,
\be\label{tilde}
\frac{\tau_{\rm tr}^2}{\tilde{\tau}^2}=\left(\frac{2\alpha-3\alpha x +7x-6}{2\alpha-8
x +6}\right)^2.\ee
Unlike the displacement contribution, Fig.~\ref{fig1}a, the photovoltaic and
quadrupole contributions are at their maximum in the absence of
the sharp component, $x=0$, and do not change parametrically in
the opposite limit: $\tilde{\tau}^{-2}|_{x=1}/\tilde{\tau}^{-2}|_{x=0}=1/4$
for $\alpha\gg 1$. However, the two contributions change significantly as a function of $x$ between $x=0$ and $x=1$ since at $x=(2\alpha-6)/(3\alpha-7)$ the
effective rate $\tilde{\tau}^{-1}$ changes sign.

Summarizing, for a sufficiently large concentration of short-range scatterers, i. e.  $x\sim1$ and $\tsh\sim\ttr$, the ratio $\ttr/\tau_*$ controlling the displacement contribution, Eqs.~\eqref{dsA} and \eqref{daA}  is of order unity (see Eq.~\eqref{star} and Fig.~\ref{fig1}a). 
For typical experimental parameters, the ratio $\tin/\ttr$ entering the inelastic contribution \eqref{dsB} is also of order unity at $T\sim 1K$  (see Sec.~\ref{ss42}). Therefore, unlike the case of smooth disorder discussed in Sec.~\ref{ss42}, in the present scenario of strong short-range disorder both mechanisms produce comparable contributions to the diagonal isotropic component $d_s$ of the photoconductivity \eqref{matrix}. At the same time, the photovoltaic and quadrupole contributions $d_{s,a}^{(D)}$, $h_s$ and $h_a$ still contain the small factor
$(\w\ttr)^{-1}\sim(\wc\ttr)^{-1}$ which is of order $\tq/\ttr$
for $\wc\tq\sim 1$. Consequently, the contributions $d_{s,a}^{(D)}$ to
the diagonal part of the photoconductivity tensor can be neglected. The terms $h_s$ and $h_a$ are also small; however, they remain the  only microwave-induced contributions to the  off-diagonal
part of the photoconductivity tensor and, therefore, are important.

Since the displacement and inelastic mechanisms are equally important in the diagonal photoresponse in the case of strong short-range disorder, $x\sim1$,
it is natural to discuss possible ways to separate their contributions experimentally. 
As we argue below, the difference in the temperature dependence of the two contributions can be 
used to extract the contributions separately. 
In Sec.~\ref{s5} the temperature dependence of various contributions is analyzed in detail.

\section{Temperature dependence of MIRO for smooth and mixed disorder}\label{s5}
\noindent 
\subsection{Temperature dependence of MIRO for disorder-broadened Landau levels}\label{ss51}\noindent
In the photoconductivity tensor, as given by
Eqs.~\eqref{matrix}, \eqref{dss} and \eqref{hss},  the only
$T$-dependent quantity is the inelastic scattering time $\tin$
entering the inelastic contribution \eqref{dsB}. According to the
analysis in Ref.~\onlinecite{DVAMP0405}, the inelastic relaxation of
the $\ve$-oscillations in the distribution function is dominated by
electron--electron (e--e) collisions. Under conditions of
the MIRO experiments, the e--e collision rate for an electron at energy
$\ve$ counted from the Fermi level $\ve_F$ is
\be\label{tauee} \frac{ 1 }{ \tau_{\rm ee}(\ve,T) }=\frac{ \ve^2 +
\pi^2 T^2}{4\pi\ve_F}\ln\frac{\kappa v_F}{{\rm
max}\{T,\wc(\wc\ttr)^{1/2}\}}, \ee
with the inverse screening length $\kappa=4\pi
e^2\nu_0$.\cite{DVAMP0405} The effective inelastic scattering time
entering \eqref{dsB} is given by the thermal average of the
out-scattering time, 
\be\label{tin} 
\tin=\{\tau_{\rm
ee}(\ve,T)\}_T\simeq 0.822\  \tau_{\rm ee}(0,T), 
\ee
where the notation
\begin{equation}\label{notation}
\{F_\ve\}_T\equiv-\int\!d\ve F_\ve\partial_\ve f_T(\ve)
\end{equation}
has been introduced.

The above results are  applicable as long as the disorder alone
controls both the momentum relaxation  and the energy dependence of
the DOS. Namely, when the contributions of the inelastic scattering
mechanisms to the momentum relaxation rate, $\ttr^{-1}$, and the quantum
scattering rate, $\tq^{-1}$, are negligible. 
This is usually the case at $T\sim1$~K, which is a typical temperature in MIRO experiments.

\subsection{Effect of interactions on the density of states}\label{ss52}\noindent
At higher
temperatures, the momentum relaxation is modified by the phonon
effects, while the DOS is influenced by both the e--e and electron--phonon interaction.\cite{zudov09,2QW,raichev08,RCS} In the limit
of overlapping LLs, the effect of the interactions on the DOS
amounts to substituting $\lambda$ in Eqs.~(\ref{curgen}) and
(\ref{f}) with
\be\label{int} 
\tilde{\lambda}(\ve,
T)=\exp\left(-\frac{\tau_{0}^{-1}+\tau_{\rm
ee}^{-1}(\ve,T)+\tau_{\rm e-ph}^{-1}(\ve,T)}{\wc/\pi}\right). 
\ee
The phonon effects will be considered elsewhere.\cite{DV} Here
we discuss the modification of the temperature dependence of the
photoconductivity tensor \eqref{matrix}, \eqref{dss}, \eqref{hss}
caused by the effect of the e--e interaction on the DOS 
(in recent experiments\cite{zudov09,2QW} the e--e interaction dominates the effect on the DOS).
As shown below, the magnitude of the above effect is controlled by the
parameter
\begin{equation}\label{gamma}
\gamma(T) = \frac{2\pi}{\wc\tau_{ee}(0,T)} \, .
\end{equation}

\subsection{Interaction-induced exponential $T$--decay of the MIRO}\label{ss53}\noindent
In order to account for the interaction-induced variation \eqref{int} of the amplitude
of the DOS oscillations (which is slow on the scale $\wc\ll T$), we make the
replacement 
\be\label{f1}
\lambda^2\equiv\{\lambda^2\}_T\to f_1(T)=\{\tilde{\lambda}^2(\ve,T)\}_T
\ee
everywhere in Eqs.~\eqref{dss} and \eqref{hss} except for the inelastic contribution $d_s^{(B)}$, Eq.~\eqref{dsB}. 
In the latter case, the replacement is, see Eq.~\eqref{tin},
\be\label{f2}
\lambda^2\frac{\tin}{\ttr}
\to f_2(T)=\left\{\tilde{\lambda}^2(\ve,T)\frac{\tau_{\rm ee}(\ve,T)}{\ttr}\right\}_T.
\ee

 At
$\gamma\ll1$, the interaction-induced
modification of the DOS is small, yielding
\bea \label{<1}
f_1&=&\lambda^2(1-\gamma/3)\,,\qquad
\gamma\ll 1\,,
\\
f_2&=&\lambda^2\left(\frac{\tin}{\ttr}-\frac{2\pi}{\wc\ttr}\right)\,,\qquad
\gamma\ll 1\,.
\label{<2}
\eea
The correction to the inelastic
contribution $d_s^{(B)}$ is temperature independent, while all other contributions
acquire a weak temperature dependence through $\gamma\propto
T^2$. 

In the opposite limit $\gamma\gg 1$,
\bea  \label{>1}
f_1&=&\lambda^2 \frac{\pi^{3/2}e^{-\gamma}
}{4\gamma^{1/2}}\,,\qquad\gamma\gg 1\,,\\\label{>2}
f_2&=&\lambda^2\frac{\pi^{5/2}e^{-\gamma}}{2\gamma^{3/2}\wc\ttr}\,,\qquad\gamma\gg 1.
\eea
In this high temperature limit, all the quantum effects
$\propto\lambda^2$ that survive the thermal averaging at $2\pi^2
T/\wc\gg 1$ are exponentially suppressed by the inelastic e--e scattering. 

Interestingly, the SdH
oscillations behave just in the opposite way: They are
exponentially suppressed by thermal smearing at $2\pi^2 T/\wc\gg 1$ but they are not
influenced\cite{maslov03,e-phSdHO,adamov06} by the interactions that lead to the dependence of $\tau_{\rm ee}^{-1}$ on $\ve$ in the form $\tau^{-1}_{\rm ee}\propto\ve^2+\pi^2T^2$.
Indeed, for such a specific form of
$\tau^{-1}_{\rm ee}$ entering Eq.~\eqref{int},
the thermal average produces the unperturbed result,
$\{\tilde{\lambda}(\ve,T)\cos(2\pi\ve/\wc)\}_T\to
\lambda\{\cos(2\pi\ve/\wc)\}_T$ for $T\gg \wc$, because in this case the
energy-dependent term in $\tau^{-1}_{\rm ee}$ effectively shrinks
the range of energy integration  and increases $\{\cos(2\pi\ve/\wc)\}_T$
to compensate $\exp(-\pi/\wc\tau_{\rm ee}(0,T))$.

\subsection{Temperature regimes of the MIRO for smooth
and sharp disorder}\label{ss54}
\begin{figure}[ht]
\centerline{
\includegraphics[width=0.95\columnwidth]{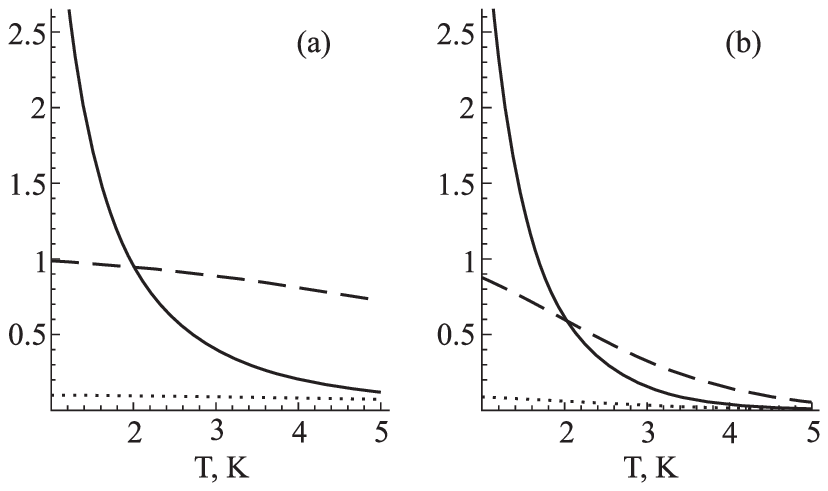}}
\caption{
{\it Solid lines:} $T$-dependence of the magnitude of the inelastic contribution \eqref{dsB}, 
$2\{\tau_{\rm ee}(\ve,T)\tilde{\lambda}^2(\ve,T)\}_T /\ttr\lambda^2$, 
for $\tin|_{T=1\,{\rm K}}=2\ttr$. 
{\it Dashed lines:} $T$-dependence of the magnitude of the displacement contribution \eqref{dsA}, $\ttr\{\tilde{\lambda}^2(\ve,T)\}_T/2\tau_*\lambda^2$, for $\ttr/2\tau_*=1$ (which corresponds to $\ttr/\tau_{\rm sh}=0.6-0.7$ in Fig.~\ref{fig1}).
{\it Dotted lines:} $T$-dependence of the magnitude of the displacement contribution \eqref{dsA}, $\ttr\{\tilde{\lambda}^2(\ve,T)\}_T/2\tau_*\lambda^2$,
 for $\ttr/2\tau_*=0.1$ [which corresponds to smooth disorder, for instance, $\ttr/\tau_{\rm sh}=0$ and  $\ttr/\tq\simeq 50$ in Fig.~\ref{fig1}, see Eq.~\eqref{star}]. 
{\it Left and right panels} correspond to (a) weak effect of interactions on the DOS at $T=1$~K$, \gamma_1\equiv 2\pi/\wc\tau_{\rm ee}(\ve_F, 1\,{\rm K})=0.01$, 
and to (b) strong effect, $\gamma_1=0.1$.
}
 \label{fig2}
 \end{figure}

\noindent
It follows from the above arguments that different temperature dependences of magnetoresistance are expected for samples with smooth
and sharp disorder.
In the former case
$\ttr/\tau_*\ll 1$, and the inelastic contribution \eqref{dsB} is bigger than all other contributions by a factor of
$\tin/\tq$, see Eq.~\eqref{BtoA}. This ratio is estimated as $\tin/\tq \sim 10-10^2$ at $T=1$~K, making the inelastic mechanism 
dominant in the whole range of temperatures where the MIRO are observed. 
The temperature dependence of the diagonal  isotropic component $d_s$ of the photoconductivity tensor \eqref{matrix} in the case of smooth disorder is then given by the
function $f_2$, Eqs.~\eqref{f2}, \eqref{<2}, and \eqref{>2}, which is shown in Fig.~\ref{fig2} by the solid line. 
Specifically, the $T^{-2}$ behavior \eqref{<2} at low $T$ crosses over to the exponential decay \eqref{>2} at high $T$.

The situation changes in the case of strong short-range component of disorder, $\ttr/\tsh\sim 1$. 
Since $\ttr/\tau_*\sim 1$ in this case, the displacement contribution \eqref{dsA} 
to the diagonal isotropic component $d_s$ of the photoconductivity tensor \eqref{matrix} 
starts to dominate over the inelastic contribution \eqref{dsB} at a certain characteristic temperature $T_1$ at which the MIRO are still strong.
The temperature $T_1$ in Kelvin is given by the relation 
\be\label{T1}
T_1=1~{\rm K}\times \frac{2}{\ttr}\sqrt{\tau_*\,\tin(1~{\rm K})}, 
\ee
see Eqs.~\eqref{BtoA}, \eqref{tin}, and \eqref{tauee}. Experimentally, $T_{1}$ is of order $1$~K for 
the case of strong sharp component of disorder since both 
$\tau_*/\ttr$ and $\tin(1~{\rm K})/\ttr$ are of order unity.
Another characteristic temperature $T_{2}$ is given by the relation $\gamma(T_{2})=1$, see Eqs.~\eqref{tin} and \eqref{gamma},
\be\label{T2}
T_2=\frac{1~{\rm K}}{\gamma^{1/2}(1~{\rm K})}\simeq 1~{\rm K}\times 0.44\sqrt{\wc\tin(1~{\rm K})}. 
\ee
It separates the temperature region $T\gtrsim T_2$ where the  LLs broadening induced by the e--e interaction becomes essential.

In the case $T_{1} \ll T_{2}$ the diagonal symmetric part $d_s$ of the phoconductivity \eqref{matrix} 
has different temperature dependence in three temperature intervals separated by  $T_{1}$  and $T_{2}$. For $T<T_{1}$ the inelastic mechanism is dominant and, as a result, 
$d_s\simeq d_s^{(B)}\propto T^{-2}$. In the interval $T_1 < T < T_2$, the inelastic contribution becomes 
smaller than the displacement contribution leading to a temperature independent magnetoresistance $d_s\simeq d_s^{(A)}\simeq \mathrm{const}(T)$. 
Finally, for $T>T_{2}$ all the contributions are suppressed exponentially, see Eq.~\eqref{<1} and Fig.~\ref{fig2}.

Clearly, the $T$--independent interval doesn't exist when $T_{2}$ is of order or smaller than $T_{1}$. This situation was probably realized in the recent experiment.\cite{zudov09}. The photoresponse is then expected to cross over from the inverse quadratic dependence $d_s \propto T^{-2}$ for $T<T_{2}$ to a much faster exponential decay  for $T>T_{2}$. 
In this case, it is difficult to separate the displacement and inelastic contributions using the temperature dependence of the MIRO only, since in the region where the two contributions have comparable magnitudes, 
their temperature dependence is almost identical (see Fig.~\ref{fig2}b). Therefore, for $T_1\gtrsim T_2$ some additional measurements aimed at determination of, for instance, the anisotropic diagonal component $d_a$ of the photoconductivity \eqref{matrix} are desirable. Since $d_a$, Eq.~\eqref{daA}, is produced solely by the displacement mechanism (the photovoltaic diagonal contributions $d_a^{(D)}$ and $d_s^{(D)}$ are negligible for $\ttr/\tsh\sim 1$), such a measurement would be a reliable tool to estimate the significance of the displacement mechanism at a given temperature. Alternatively, for a fixed polarization of the microwaves, one can analyze the photoresponse in the vicinity of ``odd nodes'' $\omega/\wc=n+1/2$, where the displacement contribution \eqref{dsA} is non-zero,
 $\sin^2(\pi\w/\wc)=1$, while the inelastic contribution $\propto\sin(2\pi\w/\wc)=0$ vanishes. 

\section{Conclusion}\label{s6}
\noindent
In conclusion, we studied the microwave-induced magnetooscillations (MIRO) in the conductivity of a 2D electron gas with mixed disorder containing both a long-range random potential and short-range scatterers. We calculated all contributions to the current that are linear in the applied dc electric field and quadratic in the microwave field. It is found that the relation between different contributions strongly depends on the strength of wide-angle scattering off disorder and the sample temperature. In general, the microwave radiation modifies both the diagonal and off-diagonal components of the conductivity tensor,  including the Hall component. Study of
anisotropy of the current may identify each contribution separately. The dominant contribution at low temperatures is the inelastic contribution, which is always diagonal and isotropic. At higher temperatures and in the presence of wide-angle scattering, the displacement mechanism may become larger than the inelastic contribution. At high temperatures the quantum corrections to the conductivity are exponentially suppressed by the  electron-electron interaction.

We believe that the possibility of measuring two types of nonequilibrium magnetooscillations---the MIRO and the HIRO---in the same sample has opened new avenues for the characterization of 2D electron systems. In particular, measuring both of them could unambiguously reveal the two-scale structure of disorder characteristic of high-mobility samples. Indeed, the relative weights of the smooth component of disorder (produced by remote impurities separated from the electron gas by a wide spacer) and the short-range component (produced by residual impurities sitting near or at the interface) manifest themselves differently in the various relaxation times that could be extracted from the experimental data. As we have demonstrated in this paper, the MIRO are sensitive to the correlation properties of disorder. Specifically, their amplitude and shape are parametrized by the inelastic relaxation time and the disorder-induced relaxation times for the zeroth, first, and second angular harmonics of the electron distribution function, with the displacement contribution to the MIRO---strongly enhanced in the presence of short-range disorder---being parametrized by the time $\tau_*$. Therefore, in addition to the routinely measured $\tau_{SdHO}$\cite{noteSdH} and $\tau_{\rm tr}$, one can: (i) extract the time $\tau_{\rm q}$ from the exponential dependence of the amplitude of the oscillations on $B$, both the MIRO and the HIRO; (ii) extract the time $\tau_*$ from the $T$ and $\omega$ dependences of the MIRO, as well as from the amplitude of the HIRO. As for the inelastic time, it can be extracted in a variety of ways: from the power-law $T$ dependence of the MIRO amplitude at lower $T$, from the MIRO and/or HIRO exponential behavior as a function of $T$ at higher $T$, as well as from the behavior of the nonlinear dc resistance as a function of the applied dc electric field.\cite{inelasticDC} Importantly, the inelastic and displacement contributions to the MIRO can then be unambiguously separated from each other. One interesting possibility worthy of experimental investigation is to extract all the relevant parameters in the absence of microwaves and use them to describe the experimental data on the MIRO without any free parameters.

\emph{Acknowledgements} We are thankful to D.N. Aristov, S.I. Dorozhkin, I.V. Gornyi, O.E. Raichev, S.A.~Studenikin, 
S.~Vitkalov, and M.A. Zudov for fruitful discussions. This work was supported by the DFG, by the DFG-CFN, 
by Rosnauka Grant No. 02.740.11.5072, by the RFBR, and by the BNL Grant No.08-002 under the Contract No. DE-AC02-98CH10886
with the U. S. Department of Energy.


\begin{thebibliography}{1}
\bibitem{noteSdH} We emphasize that the quantum relaxation time $\tau_{SdHO}$, routinely obtained from the measurements of the damping of Shubnikov-de Haas oscilations, includes the effect of macroscopic potential variations (inhomogeneous broadening of Landau levels). Such $\tau_{SdHO}$ might be much shorter than the homogeneous $\tau_q$ responsible for the damping of the MIRO. 
Therefore, damping of the MIRO or HIRO [at suficienly low temperature, see Eq.~\eqref{int}] provides a reliable method of extracting the homogeneous quantum relaxation time $\tau_q$.
\end{thebibliography}
\end{document}